# Field-Free Deterministic Writing of SOT-MTJ by Unipolar Current

Tengzhi Yang, Meiyin Yang, Lei Zhao, Jianfeng Gao, Qingyi Xiang, Wenjing Li, Feilong Luo, Li Ye and Jun Luo

*Abstract*—We propose a manufacturable solution for field-free writing of perpendicular SOT-MTJ without sacrificing integration density on a 200 mm wafer. The field-free writing operation can be achieved by unipolar current pulses via engineering the interlayer exchange coupling in SOT-MTJ thin films. The proposed device can reach a high writing speed of up to 1 ns and work properly at temperature of 100 ℃. The deterministic writing of SOT-MTJ by unipolar current offers an effective approach for high density SOT-MRAM integration.

## I. Introduction

Spin orbit torque random access memory (SOT-MRAM) has attracted great attention due to its non-volatility, high speed and infinite endurance so that it holds great advantages in future on-chip memory such as high speed cache [1-6]. To date, SOT-MRAM has made great development and is promised to achieve higher speed and lower power dissipation in the future [7]. However, SOT-MRAM still has some challenges [1, 8]. For example, typical SOT-MTJ requires an external in-plane magnetic field to realize deterministic switching, while it is nearly inapplicable for integrated circuit design. The requirement of external in-plane field during the writing process has been hampering its applications.

To solve this problem, some manufacturable solutions have been proposed. The previous work adopts in-plane SOT-MTJ structures to avoid random switching existing in perpendicular SOT-MTJ [9, 10]. Recently, field-free writing of SOT-MTJ devices with integration of Co magnets were achieved [11]. In the meanwhile, STT assisted SOT switching methods [12, 13] have been proposed which can also realize deterministic switching. Except from all methods mentioned above, there are many other approaches trying to realize deterministic switching without external field [14-19]. However, field-free writing methods are still under research for the manufacturable and high-density integration of SOT-MRAM.

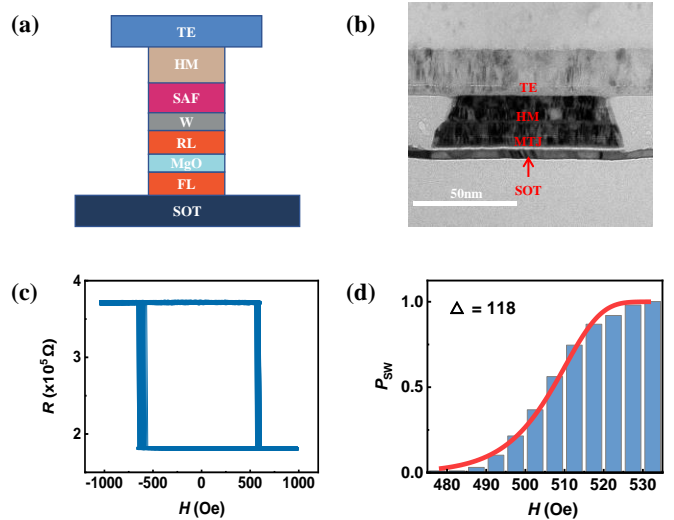

Fig. 1. (a) SOT-MTJ device schematics. (b) The cross-sectional TEM image of the unipolar writing SOT-MTJ device. (c) *R-H* minor loops of the proposed SOT-MTJ measured by 100 times. (d) Switching probabilities dependence on the magnetic field.

In this work, we propose a field-free writing solution without sacrificing integration density for perpendicular SOT-MTJs on a 200 mm CMOS compatible platform. In this work, the deterministic writing was achieved by engineering the interlayer coupling between free layer (FL) and reference layer (RL) through MgO. Heat induced by current pulses adjusts the coupling strength, resulting in parallel (P) or antiparallel (AP) states of SOT-MTJ by unipolar current pulses. The proposed SOT-MTJ shows high speed, and is able to work properly under the maximum working temperature of typical chips at 100 °C, which is suitable for high density and large retention cache memory for industrial chip design.

## II. Experiments

The SOT-MTJ device in this study using top-pinned W(5 nm)/CoFeB(FL)/MgO(1.2 nm)/CoFeB(RL)/SAF film stack structure with perpendicular magnetic anisotropy (PMA) is shown in Fig. 1(a). Interlayer coupling between FL and RL of the MTJ in this work was engineered by adjusting the thickness of MgO based on our previous work [20]. The magnetic film was grown by magnetron-controlled sputter with the pressure of $9\times10^{-7}$ Pa. PMA of the SOT-MTJ thin film is maintained after annealing in vacuum at 350 °C for 1 hour. The device was fabricated by CMOS compatible technology on our 200 mm

This work was supported by Chinese Academy of Sciences (Grant Nos. XDA18000000, Y201926 and 2020118) and the National Key R&D Program of China (2018YFB0407602). The authors also thank Hisilicon Technologies for financial support.

T. Yang, M. Yang, L. Zhao, J. Gao and J. Luo are with Key Laboratory of Microelectronic Devices and Integrated Technology, Institute of Microelectronics, Chinese Academy of Sciences (IMECAS), Beijing 100029, China. T. Yang and M. Yang contribute equally to this work. (e-mail: luojun@ime.ac.cn, yangmeiyin@ime.ac.cn).

T. Yang, M. Yang L. Zhao, and J. Luo are with University of Chinese Academy of Sciences (UCAS), Beijing 100049, China.

Q. Xiang, W. Li, F. Luo and L. Ye are with Hisilicon (Shanghai) Technologies Co., Limited (e-mail: yeli3@hisilicon.com).

manufacturable platform. The MTJ and bottom SOT electrode were patterned by deep ultraviolet lithography (DUV) and etched by inductively coupled plasma etching (ICP). We adopt high resolution transmission electron microscope to analyze cross-section morphology of the device. The electric transport measurements are conducted using magnetic field probe station by East Changing Technologies, China. The pulse duration of the injected current pulse ($t_{pulse}$) is set to 50 ns, except being indicated otherwise.

## III. RESULTS AND DISCUSSIONS

### A. Basic magnetic properties

The fabricated MTJ device with CD of ~140 nm is presented in the TEM image in Fig. 1(b), sitting on a bottom SOT electrode with the width of 300 nm. TMR value of 111 % is extracted from the loops of resistance ($R$) vs. magnetic field ($H$) measured by 100 times, in which the offset field is shifted to zero, shown in Fig. 1(c). Then, the switching possibility ($P_{SW}$) dependence on the $H$ obtained in Fig. 1(c) are plotted in Fig. 1(d), which can be described by the equation below [21]:

$$P_{SW} = 1 - \exp\left[-\left(\frac{H_k f_0 \sqrt{\pi}}{2R\sqrt{\Delta}}\right) erfc\left[\sqrt{\Delta}\left(1 - \frac{H}{H_k}\right)\right]\right] \quad (1)$$

where $f_0$ is attempt frequency, $H_k$ is anisotropy field and $erfc$ is complementary error function. By fitting $P_{SW}$-$H$ curve based on the equation (1), the thermal factor ($\Delta$) of our devices is estimated to be as large as 118, indicating that the device possesses very high retention. During all writing tests, no offset field is exerted on FL by applying a perpendicular magnetic field, which can be avoided by optimizing the SAF layer.

### B. Unipolar Deterministic Switching

$R$ vs. current density ($J$) curves with a continuously-changed current pulse without in-plane external field is shown in Fig. 2(a) and (b), where the magnetization of RL was preset to -$z$ direction in Fig. 2(a) and to +$z$ direction in Fig. 2(b). Regardless of the magnetization of RL, the switching loops are similar, where large current pulses always set the MTJ to AP states. Decreasing the amplitude of the current pulse, the SOT-MTJ is switched to P states. To reveal the mechanism of this phenomenon, we show the relationship between $R$ and $H$ in Fig. 2(c) measured under different pulse amplitudes using continuous pulse sequence and the pulse frequency is identical to that employed in $R$ vs. $J$ curves. The derived critical switching magnetic field ($H_c$) and offset field are shown in Fig. 2(e). The unipolar switching process have two periods. Firstly, when $J$ is increased from 0 to 6.5×10$^7$ A/cm$^2$, the offset field significantly rises and the $H_c$ reduces in the meanwhile, indicating that the parallel state is getting more favorable. This phenomenon is due to the enhancement of ferromagnetic interlayer exchange coupling by the current pulses. As we have discussed in the previous study [20], there is orange peel type exchange coupling existing between FL and RL. The orange peel coupling results in ferromagnetic coupling of FL and RL in small PMA system [21]. When increasing the writing pulse density from 0 to 6.5×10$^7$ A/cm$^2$, the heat induced by current pulses reduces PMA of FL, which enhances the ferromagnetic coupling of FL and RL via orange peel coupling, leading to the

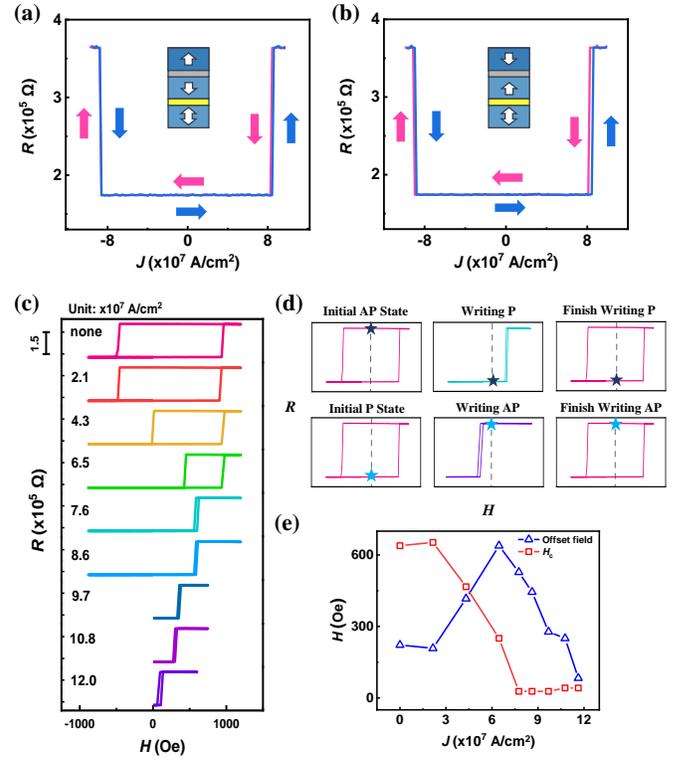

Fig.2. The $R$ of the SOT-MTJ changed with the sweeping of $J$ without external magnetic field when RL was preset to (a) -z and (b) +z direction. (c) $R$ measured by sweeping perpendicular magnetic field $H$ when injecting writing pulses under different current densities. (d) The AP-to-P (dark blue star) or P-to-AP (light blue star) behaviors driven by unipolar currents can be explained by the magnetic dynamics. (e) $H_c$ and offset fields as a function of writing current densities.

P state after current injection. The second period begins when $J$ increases from 6.5×10$^7$ A/cm$^2$ that the offset field starts to decrease. The phenomenon is derived from the loss of PMA for FL by the heating during the current injection, which weakens the ferromagnetic coupling in perpendicular direction [22]. The ferromagnetic coupling quickly disappears at a higher temperature, which leads to a stable AP state of MTJ due to the dipole field from the SAF with the pulse over 1.1×10$^8$ A/cm$^2$ (see the light blue star in Fig. 2(d)). Because of the offset change and $H_c$ reduction by SOT, stable P and AP states of MTJ were obtained after pulses around 6.5×10$^7$ A/cm$^2$ and 1.1×10$^8$ A/cm$^2$ respectively as illustrated with the light blue star and dark blue star in Fig. 2(d).

To investigate the SOT effect on the switching of the SOT-MTJ, we measured the MTJ resistance versus writing current density under external in-plane magnetic field of ±500 Oe. The $R$-$J$ loop shown in the Fig. 3(a) and (b) shows typical chiral switching behavior by SOT, where the deterministic orientation of FL driven by fixed current pulses depends on the direction of the in-plane field, which is consistent with the switching mechanisms under SOT effect [23]. The chiral switching indicates the effect of SOT contributes greatly to the switching of the SOT-MTJ. Thus, the switching of the SOT-MTJ by unipolar currents in the absence of in-plane field is primarily due to the SOT effect rather than heating, while the orientation of the switching is determined via the modulation of interlayer couplings by temperature.

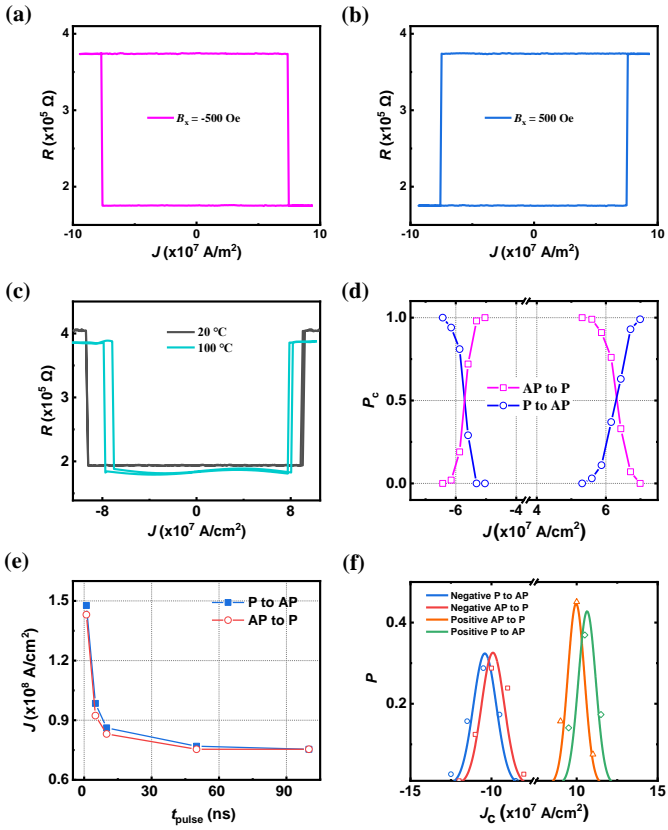

Fig.3. The current-induced magnetic switching of the proposed SOT-MTJ with external in-plane field of (a) -500 Oe and (b) 500 Oe. (c) The measured resistance R of the SOT-MTJ by sweeping current pulses without external in-plane field for temperature of 20 °C and 100 °C, respectively. (d) Switching probabilities at negative current and positive current pulses. (e) $J_c$ vs. $t_{pulse}$ without in-plane magnetic field. (f) The distribution (P) of $J_c$ without in-plane magnetic field.

Because the deterministic writing of our SOT-MTJ is influenced by temperature, whether the device can work on the extreme environment will be a question to the practical application. The device should work at the temperature up to 85 °C for the requirement of industrial chip standard. The temperature response of the switching behaviors was measured, as shown in Fig. 3(c). The rising temperature only brings with a smaller switching current density ($J_c$) and reduced TMR, because of the direct temperature dependence on spin polarization according to the Jullière's model [24, 25]. The $J_c$ decreases by up to 9 %, which can still ensure the adequate writing margin. The writing of the SOT-MTJ by unipolar currents maintains at 100 °C. The measured result illustrates that the device can maintain stable electrical performance. Therefore, the proposed device can satisfy the standard of industrial level chip design.

The dependence of current-induced switching probability ($P_c$) from AP (P) to P (AP) in Fig. 3(d) indicates that both positive and negative current pulses can lead to deterministic writing for the SOT-MTJ. In the meanwhile, the deterministic switching behavior with unipolar writing current is also valid when the writing speed is as fast as 1 ns (in Fig. 3(e)). In addition, we sampled more than 50 devices in random from two different wafers. The distribution of $J_c$ shown in Fig. 3(f) is tight and obeys a gaussian distribution, which indicates the good device uniformity in the 200 mm wafer.

## IV. CONCLUSION

In summary, the field-free deterministic writing of SOT-MTJ device with unipolar switching mode has been proposed. The deterministic writing is achieved by the modulation of ferromagnetic coupling between FL and RL due to the heating of current pulses. Furthermore, the proposed device can reach up to 1 ns per writing operation and can work properly under the temperature of 100 °C, which can meet with high density SOT-MRAM integration of industrial chip standard.